\documentclass[a4paper,preprint]{elsarticle}

\makeatletter
\def\ps@pprintTitle{%
  \let\@oddhead\@empty
  \let\@evenhead\@empty
  \def\@oddfoot{\reset@font\hfil\thepage\hfil}
  \let\@evenfoot\@oddfoot
}
\makeatother

\usepackage{float}
\usepackage{latexsym}
\usepackage{graphicx}
\usepackage{amsmath} 
\usepackage{amsfonts} 
\usepackage{color}
\usepackage{tikz}
\usetikzlibrary{automata,arrows,positioning,calc, arrows.meta}
\usepackage{array}
\usepackage{booktabs}

\usepackage{amssymb,amsmath}
\usepackage{graphics,graphicx}
\graphicspath{{./images/}}

\DeclareGraphicsExtensions{.pdf,.eps}

\usepackage[english]{babel}
\usepackage{bbm,theorem}
\usepackage{a4wide}

{\theorembodyfont{\upshape}

}

\newcommand{\etal}{{\it\,et al.~}}

\usepackage{xcolor}

\usepackage{setspace}
\title{A rich life cycle model of labor supply in Finland}
\author[Tela]{Antti J. Tanskanen}
\ead{antti.tanskanen@ek.fi}
\address[Tela]{The Finnish Pension Alliance TELA, Salomonkatu 17 B, FI-00100 Helsinki, Finland.\\e-mail: antti.tanskanen@gmail.com}
\date{\today}
\begin{document} 
\selectlanguage{english}
\begin{frontmatter}

\date{\today} 

\begin{abstract}
A life cycle model of consumption and labor supply describes employment decisions of a collection of individuals during their lifetime. We develop a life cycle model describing a heterogeneous population operating in Finland under a wide variety of employment states and life situations. A rich life cycle model requires a large state space representing the possible states of simulated agents. The results demonstrate that the model reproduces a number of statistics  of the Finnish employment market such as the age structures of employment rate and unemployment rate, distributions of observed effective marginal tax rates and participating tax rates, and proportion of part time work. As an application of analysis of a reform, we analyze how the program of Orpo government influences employment and public finances in Finland. 
\end{abstract}
 
\end{frontmatter}

\section{Introduction}

A life cycle model of consumption and labor supply (abbreviated LCM in the following) describes how people make decisions on whether to work or not, and how much to work. If saving is included in an LCM, it can be used to describe how people smoothen consumption over their life-times (Modigliani and Brumberg 1954). 

Stochastic LCMs have been applied to assess the influence of a social security reform on employment and unemployment rates (Hakola and Määttänen 2007, Määttänen 2014, Tanskanen ja Kotamäki 2021), retirement (Jiménez‐Martín and Sánchez Martín 2007, Blau 2008, Rust 1989),
and on the internal rate of return on pension contributions (Tanskanen 2020b). 
An LCM is often used to study optimal savings behavior during an individual's life cycle (Modigliani and Brumberg 1954) and consumption patterns (Gourinchas and Parker 2002).
LCMs have demonstrated more successes than failures in these contexts (Browning and Crossley 2001).

\subsection{Analyzing social security reform}
A social security reform often aims either to improve social security of certain groups of individuals or to revise the benefits to encourage employment. 
In both cases, estimating how the reform affects the financial sustainability of a social security scheme is crucial for the viability of the reform.

Due to complexity of social security schemes, the influence of a social security reform on employment and financial sustainability is quite hard to analyze, especially when behavior of people is taken into account. If a reform is not financially sustainable, there is little reason to adopt it. 

A stochastic LCM of labor supply simulates how rational individuals make employment decisions, when the future is uncertain. 
An individual makes her employment decisions by maximizing the expected utility in an LCM. The cornerstone of this modeling approach is the utility function, which balances the value of free time against the benefits of consumption. Consumption, facilitated by wages, social security benefits, and savings, must be weighed against the opportunity cost of lost free time due to work. Striking this balance determines the labor supply.
An LCM aims to provide insight into how individuals respond to policy reforms, such as changes in social security or taxation schemes.


\subsection{Population, employment decisions and state transitions}
Work incentives vary across different household types. A single household with children, a couple with children, a couple without children, and a single household without children all exhibit different optimal behaviors. A comprehensive LCM should be able to find the optimal behavior for each of these groups at various stages of life, taking also into account that in the future, the single agent may be part of a couple.

Transitions between employment states are influenced by a mix of factors. Some transitions, such as changing employment status, are deliberate decisions made by the individual. Others, like moving to disability pension, occur with exogenous probabilities. Certain transitions, like quitting a job, may result from either an individual’s decision or external circumstances. Agents decide their next course of action, such as remaining employed or transitioning to unemployment, taking all these possibilities into account.

\subsection{Finding optimal behavior}
Dynamic programming is a well-known method for solving an LCM (see, e.g., Heer and Maussner 2009,  Rust 1989, Määttänen 2014) giving a good approximation to the optimal behavior. The approximation converges toward the globally optimal solution in the limit of infinitely tight grid (Puterman 1990). However, applicability of dynamic programming is constrained by its large demand for resources and finite available computation time (Sutton and Barto 2018). Dynamic programming finds optimal behavior best in a relatively small state space with less than 10 discrete or discretized state variables. 

Modeling a modern social security scheme in, e.g., a Nordic country, requires a large number of state variables to describe the social security benefits and wages in detail. In an LCM describing Finnish social security (Tanskanen 2020a), the number of state variables was 13. To find an optimal behavior in this model, dynamic programming would need to solve the model on a grid with $~10^{21}$ states, which is computationally unfeasible. Solving an LCM in a large state space requires the use of approximative algorithms (Rust 1997).  

Recent advances in machine learning algorithms enable solving a rich LCM with a large state space approximately. 
Reinforced learning (abbr. RL) algorithms are alternatives for dynamic programming that can find an approximately optimal solution even for an LCM with a huge state space. 

Although theoretical knowledge of convergence properties of RL algorithms is not fully developed, the performance of RL algorithms, such as ACKTR (Wu~\etal 2017),  seems to be good in practice (Tanskanen 2022). This is supported by theoretical results on the convergence (Wang~\etal 2019). 
When the state space is huge, the algorithm must be able to generalize from the observed states to the unobserved ones. This makes deep reinforced learning methods based on neural networks well suited for this task. 

\subsection{Comparison with other methods}
The most common way to estimate the impact of a reform on employment is to use an appropriate elasticity describing the response of employment on the examined change (Cahuc \etal 2014). This  is a solid, experiment-based way of estimation. 
For example, if a reform changes the maximum duration of unemployment benefit, the elasticity of the unemployment duration to the potential duration of unemployment benefit is the appropriate elasticity for the analysis of the response. 
Ministry of Finance's (2024) estimates of the impact of reforms are mostly made using appropriate elasticities.

Each employment supply elasticity describes a single pathway influencing a particular group of people. For example, elasticity of unemployment duration with respect to potential duration of unemployment benefit will only measure how the length of unemployment spells will respond to a change in the maximum duration of unemployment benefit. To estimate the response of unemployment to a change in the unemployment benefit level, we need a different elasticity. 
This raises important questions: How should various responses be aggregated? Are the pathways correlated? Are the same individuals affected via multiple pathways? Can the responses be summed without double-counting? These are not issues with an LCM.

Microsimulation is used to estimate how demographic and policy changes influence individual outcomes, and it enables a better understanding of the effects of chosen policies. Microsimulation is often based on register data on, e.g., wages and social security benefits. 
A microsimulation model provides a static analysis of how a reform influences net income across the population, without accounting for dynamic employment decisions. 
For example, Sisu microsimulation model computes static estimates of benefit expenditure and various statistics based on comprehensive Finnish register data (Statistics Finland 2018). 

By incorporating labor elasticities, a microsimulation model can factor in behavioral responses to policy changes (Kotamäki~\etal 2018). In this way, a microsimulation model can estimate the impact of a reform on public finances and employment outcomes. While a microsimulation model focuses on specific pathways, an LCM considers a broader range of dynamical of mechanisms through which reforms affect employment decisions. 
Consequently, an LCM is likely to predict a greater impact of a policy reform on employment compared to a microsimulation model incoporating labor elasticities.

A dynamic general equilibrium model (DGE) provides a more general picture of economy's response to a policy reform than an LCM. A DGE describes how the reform influences national economy demand, which yields a more detailed picture of the economy (Heer and Maussner 2009). 
While a DGE provides a more general analysis of employment and public finances than an LCM, a DGE requires a large amount of parameters and mechanisms.
Examples of DGEs used Finland include Aino (Silvo and Verona 2020, Kortelainen and Ripatti 2010) and FOG (Alho~\etal 2023), 

Compared to the DGEs used in Finland, the LCM considered here provides a more detailed picture of how individuals make their employment choices in the presence of detailed description of social security. 
Compared to the use of elasticities or to an extended microsimulation, an LCM includes more influence pathways which enables a more detailed analysis of a reform. 
We believe that an LCM is a good compromise between generality and complexity. 

\subsection{This study}
In this study, our primary objective is to develop a detailed life cycle model describing Finnish social security benefits and contributions, taxation, and employment.
The model focuses on agents’ decisions whether to participate in the labor market and, if so, how many hours to work. 
The LCM is calibrated to replicate observed labor market outcomes and benefit expenditures as accurately as possible.

The model builds upon the earlier work of Tanskanen (2020a) and Tanskanen and Kotamäki (2021), which themselves were based on the contributions of Määttänen (2014) and Määttänen and Hakola (2007). 
By solving a robust, well-calibrated LCM, this study aims to provide tools to get insight into impacts of various social security and tax reform.

The study is organized as follows.
Chapter 2 introduces a detailed life-cycle model of labor supply and consumption, capturing individual and household dynamics. 
 Chapter 3 outlines the methodology of solving the LCM, covering model calibration and data sources. Chapter 4 presents the baseline results and compares them to observed data. Chapter 5 applies the model to evaluate the social security reform of Orpo government. Finally, Chapter 6 discusses the findings and their broader implications.
 
\section{Life cycle model of labor supply}

An LCM describes how an individual makes optimal choices on whether to work or not, and how much to work. An informed choice requires a detailed description of benefits, wages, taxes and social security contributions during various life stages. 

\subsection{Heterogeneous population}
To describe an actual population, the LCM must be able to capture the behavior of a heterogenic collection of people in a wide variety of life situations.
A large, diverse simulated population also helps ensure that various choices occur with realistic frequencies, while heterogeneity captures the variety of circumstances influencing individual behavior. 

The analyzed population should be a realistic representation of the Finnish population, capturing the key demographic and other characteristics. 
We call simulated individuals agents to make a clear distinction to real individuals. The simulated population of agents is based on the Finnish population structure (Statistics Finland 2024e). 
The typical size of population in a simulation is in range $50,000 - 100,000$ chosen to correspond to cohort sizes in Finland and to be sufficiently large to keep statistical noise at a low, realistic level.

\begin{table}[ht]
\centering
\begin{tabular}{|l|c|c| } 
 \hline
Parameter & Symbol & Values\\
 \hline
 	age at time $t$ & $a_t$ & 18-100 years\\
        age at the beginning & $a_0$ & 18 years\\
        socioeconomic group & $g$ & 1-6\\
 	size of population & $n_{pop}$ & 1-100,000\\
        retirement age & $r_{age}$ & 63-68 years\\
        insurance obligation ends & $r_{max}$ & 68-70 years\\
        discount factor & $\gamma$ & 0.92\\
        time step & $\Delta t$ & 0.25 years\\
 \hline
\end{tabular}
 \caption{Parameters of simulation}
\label{table:parameters}
\end{table}

Each agent belongs to one of three socioeconomic groups and is either male or female.  
Socioeconomic groups correspond to low, middle, and high income levels. In 2018, the relative sizes of these groups is were for men 28\%, 34\%, and 28\%, while for women, the relative sizes of the groups were 25\%, 38\%, and 37\%. Over time, the proportion of individuals in the higher income group has increased slightly. By 2024, this group is projected to comprise 30\% of men and 40\% of women (Statistics Finland 2024e).

Each agent in either part of a couple or single. Relationships are formed and dissolved based on externally estimated probabilities derived from statistics on Changes in Marital Status (Statistics Finland 2024b). Transition probabilities take into account hypogamy and hypergamy in the partnerships (Mäenpää 2015). For simplicity, each agent can form a household with a single agent of the opposite sex. 

Agents can have children either as part of a couple or while single. 
The probability of having a child does not depend on how many children a family already has. There are both single-parent families and two-parent families in the model. 
Children are tracked until they reach the age of 18 years, after which they are removed from the simulation. 

Life trajectory of each simulated individual is followed from age 18 years until age 100 years. 
Each agent makes employment decisions on a quarterly basis from age 18 to age 75. Beyond age 75, a static simulation during which the employment state remains unchanged is conducted up to age 100. 
Initially, individuals are distributed across employment states—such as student, full time employed, unemployed, and part time employed—according to Employment Statistics (Statistics Finland 2024a). 
No individual-level register data is used, even though the framework allows for its inclusion. The aggregate results are scaled to represent Finland's population.

\subsection{Utility function}

The goal of each agent is to maximize her or his utility that describes how the agent values consumption, free time and time with children. 
To assess how good a decision is, an agent estimates the expected discounted future utility

\begin{equation}\label{eq:utility}
E\sum_{t=t_0}^{T}\gamma^tu(a_t,c_t,s_t,g,h_t),
\end{equation}
where $a_t$ is agent's age at time t, $c_t$ is consumption, $s_t$ is the employment state and $h_t$ is weekly work hours. 
Future cash flows are not as valuable as instant cash flows, which requires the presence of discounting factor $\gamma$ reducing the importance of future cashflows.
Since saving is not included, consumption $c_t$ equals net income $n_t$ minus VAT. 

As in Määttänen (2014), utility function for one-period utility is the logarithmic utility function with certain extra terms
\begin{equation}\label{eq:logutility}
u(a,c,s,g,h)=\begin{cases}
\log(c/D)+\kappa_{s,g,h}-\mu_{a,s,g,h}, \,\text{if the agent is alive,}\\
0, \text{else,}
\end{cases}
\end{equation}
where penalty term $\kappa$ describes the value of the lost free time.
Term  $\mu$ describes how the value of free time increases near the lowest retirement age. 
Consumption $c_t$ is adjusted by wage inflation $D$ to keep the ratio of the value of free time and the value of wage constant.
We assume that couples divide their consumption equally, and compute the utility functions accordingly. 

\subsubsection{Penalty terms}

Penalty term $\kappa$ in Equation (\ref{eq:logutility}) describes the value of lost free time. 
Values of $\kappa$ in Table \ref{table:kappa_values} were calibrates so that the predicted employment rate 
corresponds to the observed employment rate in Employment statistics (Statistics Finland 2023). 

The less $\kappa$ penalizes, the higher is the implied participation elasticity.
For women, $\kappa$ penalizes less than for men. 
This is consistent with Kotamäki (2014), who shows that participation elasticity of women is typically higher than that of men. 

\begin{table}
\centering
\begin{tabular}{ |l|c|c| } 
 \hline
Parameter & Men & Women\\
 \hline
        $S_{age}$ & $r_{age}- 5 y$ & $r_{age} - 3 y$\\
        $S_{ret}$ & $r_{age} + 15 y$ & $r_{age} + 15 y$ \\
        $q_1^{s,h}$ & 0.075h/40 & 0.065h/40\\
        $q_2^{s,h}$ & 0.035h/40 & 0.015h/40\\
\hline        
        $\kappa$ child home care allowance & 0.005 & 0.050\\
        $\kappa$ under 3y child & 0.005 & 0.010\\
        $\kappa$ student & 0.000 & 0.000\\        
        $\kappa$ retirement & 0.000 & 0.000\\
        $\kappa$ sick leave & -0.500 & -0.500\\
        $\kappa$ unemp young & -0.250 & -0.100\\
        $\kappa$ unemp middle & -0.150 & -0.400\\
        $\kappa$ unemp elderly & -0.100 & -0.100\\
        $\kappa$ work 8 h per week & -0.360 & -0.270\\
        $\kappa$ work 16 h per week & -0.390 & -0.320\\
        $\kappa$ work 24 h per week & -0.450 & -0.345\\
        $\kappa$ work 32 h per week& -0.550 & -0.365\\
        $\kappa$ work 40 h per week & -0.705 & -0.490\\
        $\kappa$ work 48 h per week & -1.400 & -1.400\\
\hline
\end{tabular}
\caption{Comparison of Women's and Men's $\kappa$ penalty terms. Variable $h$ denotes weekly work hours.}
\label{table:kappa_values}
\end{table}

Penalty term $\kappa$ depends on employment state $s$. For full time and part time work, $\kappa$ also depends on weekly hours worked $h$. 
For the unemployment states, $\kappa$ is the penalty associated with being unemployed if the employee has decided to quit work. If the employee has been laid off, $\kappa$ equals zero for the unemployment states.
For child home care allowance state, penalty $\kappa$ encourages transition to the state since it increases time with small children. 

A retiree would prefer not to work full time (Polvinen\etal 2023). Term
\begin{align}
\mu_{a,s,g,h} &= q^1_{s,g,h}\max(0,\min(a,r_{age}) - S_{age}^g) + q^2_{s,g,h}\max(0,\min(a,S_{ret}) - r_{age}),
\label{eq:mu}
\end{align}
represents reduced preference to work just before and after the lowest retirement age.  This has the consequence that an agent will retire close to the lowest retirement age.
Without additional preference $\mu$ for free time, working during retirement would be more common in the simulation than observed. 

%
%
\subsection{State space}
The state space describes all possible states of two agents and their children. 
Since the two agents may or may not form a couple, a state vector 
describes either a household consisting of two persons and their children; or two separate single households and their common children. 
In this way, we can simulate marriage dynamics and include children in the LCM. 

State space is 67 dimensional counting only actual state variables. 
In the computational implementation, there are a two indicator variables and one-hot encoding of discrete state variable. These are not strictly necessary, but will improve convergence of the RL algorithm. 
The variables describing the state of an agent are shown in Table \ref{table:statevariables}.

\begin{table}[h!]
\centering
\renewcommand{\arraystretch}{1.2} 
\begin{tabular}{|l|l|l|l|l|}
\hline
Common States & Employment & Retirement & Unemployment & Time-to \\ \hline
children $<3 y$ & employment state & early-ret. paid & employment condition & until disability\\
children $<7 y$ & work hours & pension accrual & emp. condition $58y$ & until student\\ 
children $<18 y$ & previous wage & pension paid & new emp. condition & until outsider\\
has a spouse & next wage & basic pension & ER unemp began & life left\\
until birth & wage basis & group &ER fund member& \\
until marriage & pink slip & time in state & ER benefit left& \\
until divorce & career length & &used ER benefit & \\
age & wage reduction& &unempwage basis & \\
fund member & paid wage& & unemp wage& \\
& & & unemp $>$ ret.age& \\
\hline
\end{tabular}
\caption{State vector variables. States in column "Common States" are common to both agents, all the other columns represent a single agent and hence are presented twice in the state vector.}
\label{table:statevariables}
\end{table}

Certain state vector components are computed based on the history of the agent. For example, employment condition depends on employment state and wages for the preceding 9 quarters. The result of this computation is updated to the state vector before the next time-step. Since the next state vector cannot be computed based only on the current state vector, the LCM is partially visible (Sutton and Barto 2018).

\subsubsection{Employment states}
There are 16 employment states in the simulation. There are two types of work states (full time work and part time work) and three types of unemployment states (earnings-related unemployed, basic unemployment, and the extended earnings-related unemployment). Four types of pensioners are present (disability pensioners, non-working retirees, full time working retirees, and part time working retirees), three types of parental leaves (mother's leave, father's leave and child home care allowance), and there are states for a long sick leave, for studying/military service, and for those outside work force.
Finally, there is a state for the deceased, which is required, since an agent may be eligible for survivor's pension. For simplicity, the model does not include a separate category for entrepreneurs.

\begin{figure}
\centering
\begin{tikzpicture}[->, >=stealth', auto, semithick, node distance=3cm]
\tikzstyle{every state}=[fill=white,draw=black,thick,text=black,scale=1]
\node[state]    (A)   {Student};
\node[state]    (B)[right of=A]                     {Full time};
\node[state]    (C)[above of=B]  {part time};
\node[state]    (D)[right of=B]   {Retired};
\node[state]    (E)[below of=B]   {Unemployed};
\node[state]    (F)[above of=D]   {Ret.+work};
\path
(A) edge[right]    (B)
      edge[right]    (C)
      edge[right]    (E)      
      edge[bend right] (D)
(B) edge[left] (A)
      edge[right] (C)
      edge[right] (D)
      edge[right] (E)
(C) 	edge[right] (A)
	edge[right] (B)
	edge[right] (D)
	edge[bend right] (E)
(D)	edge[right]     (F)
(E) 	edge[left]                 (B)
 	edge[left]                 (D)
	edge[bend left] (C)
(F) 	edge[left]                 (D);	

\draw[dashed, rounded corners=10pt, color=gray, very thick] (1.25,-4.5) rectangle (4.55,4.6);
\node[align=left, font=\sffamily\scriptsize] (workforce) at (3, 4.2){Work force};
\draw[dashed, rounded corners=10pt, color=gray, very thick] (4.8,-1.5) rectangle (7.15,4.6);
\node[align=left, font=\sffamily\scriptsize] (workforce) at (5.6, 4.2){Retired};

\end{tikzpicture}
\begin{tikzpicture}[->, >=stealth', auto, semithick, node distance=3cm]
\tikzstyle{every state}=[fill=white,draw=black,thick,text=black,scale=1]
\node[state]    (B)[right of=D]  {Work force};
\node[state]    (D)[above left of=B]   {Sick\ leave};
\node[state]    (C)[right of=D]   {Disabled};
\node[state]    (E)[right of=B]   {Retired};
\node[state]    (G)[below left of=B]   {Out of wf};
\path
(B) edge[right]     (D)
      edge[right]     (E)
	edge[left]                 (G)
	edge (C)
(C) edge[bend left] (E)
(D) 	edge[right]     (C)
	edge[right]     (B)
(G) 	edge[right]                 (B)
      edge[bend right] (E);
\end{tikzpicture}
\caption{Transitions between employment states. Possible transitions between work force, retirement, out of work force, sick leave and disability pension.}	
\label{fig:fig3}
\end{figure}
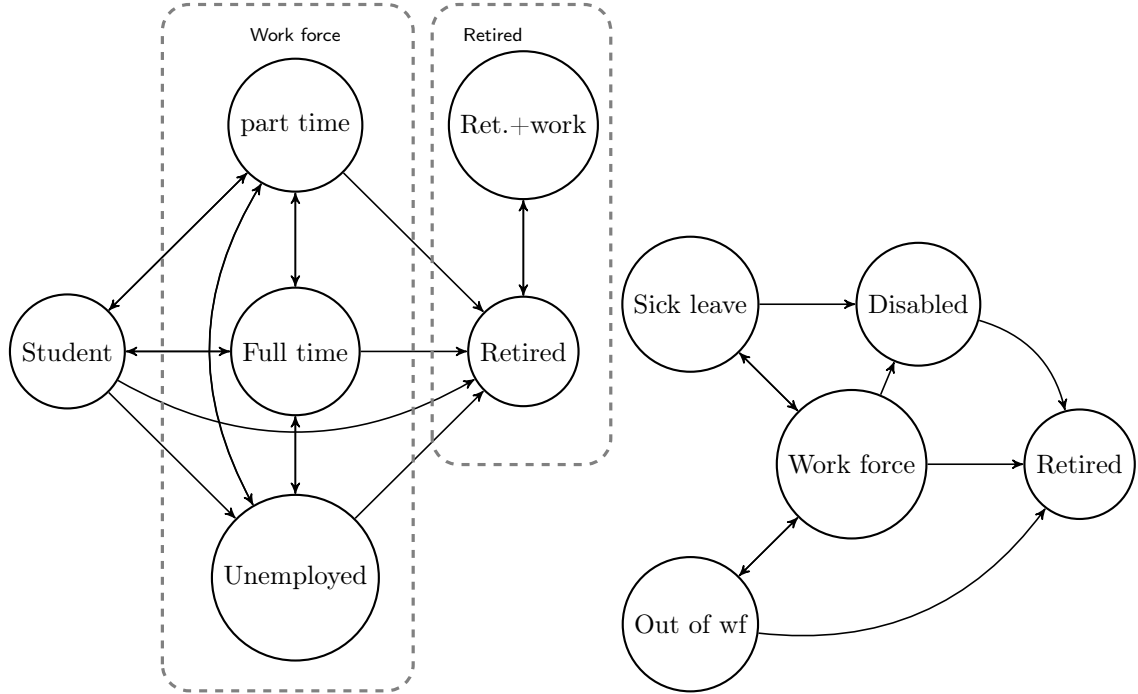

\subsubsection{Transitions between employment states}

Each agent chooses whether to work, how much to work and when to retire. No other decisions are made by the agent. 
Transitions between employment states are possible four times per year. 

Table \ref{table:table_decision} shows which employment state transitions are decisions made by an agent and which are not.
For example, searching for an employment, for a new job, or resigning from a job are a decision. 
Certain transitions occur at a prescribed probability. For example, having a child at any time-step occurs at an exogenic probability (Table \ref{table:table_decision}).
In addition to transitions shown in Table \ref{table:table_decision}, death is due to a chance in the LCM.

\begin{table}
\centering
\begin{tabular}{ |l|llllllllllllll| p{5cm}} 
 \hline
From / To & FT & Re & Di & Un & Mo & Fa & Su & RP & RF & PT & Ou & St & Lm & Si \\
 \hline
        Full time work & D & D & E & E & E & E & D & - & - & D & E & E & D & E\\
        Retired & - & D & E & - & - & - & - & D & D & - & - & - & - & - \\
        Disabled & - & E & E & - & - & - & - & - & - & - & - & - & - & -\\
        Unemployed & D & D & E & E & E & E & D & -  & - & D & E & E & D & E \\
        Mother's leave   & D* & D* & E & E & E & E & D* & - & - & D* & E & E & D* & E \\
        Father's leave    & D* & D* & E & E & E & E & D* & - & - & D* & E & E & D* & E \\
        Child home care all.  & D & - & E & E & E & E & D & - & - & D & E & E & D & E \\
        Retired \& PT & - & D & E & - & - & - & - & D & D & - & - & - & - & - \\
        Retired \& FT & - & D & E & - & - & - & - & D & D & - & - & - & - & - \\
        Part time work     & D & D & E & E & E & E & D & - & - & D & E & E & E & E \\
        Outside of WF     & - & - & E & E & E & E & E & - & - & D & E & E & E & E \\
        Student               & -  & - & E & - & E & E & E & - & - & D & E & E & E & - \\
        Labor mkt supp. & D & D & E & - & E & E & E & - & - & D & E & E & D & E \\
        Sick leave           & D\^ & E & E & D\^ & E & E & D\^ & - & - & D\^ & - & - & D\^ & E \\
\hline
\end{tabular}
 \caption{Transitions between employment states. Here D denotes "decision", E denotes "Exogenic probability", "-" denotes not applicable, D* denotes decision after external probabilities change employment state and D\^\  denotes decision after external probabilities change employment state if not disabled. FT stands for full time, PT for part time and WF for work force.}
\label{table:table_decision}
\end{table}

In addition to transitions within work force, there are transitions between work force, sick leave and outside work force. 
Transitions from work force states to sick leave occur with a probability that depends on group and employment state. The duration of the sick leave is random, and some of those that stay in sick leave for one year are moved to disability pension.  The probabilities are defined in such a way that the observed proportion of population in disability pension is reproduced in the model.
Except for retirement, transitions to outside work force occur randomly. They are calibrated so that size of work force corresponds to observation. Similarly, transitions to mother's leave and father's leave occur randomly.
After a sick leave, a parent's leave or a stay outside work force, an agent moves back to work force and makes decision whether to work or not.
To make the optimization problem more feasible, each agent knows the time until the next random state transition.

\subsection{Employment}
\subsubsection{Work hours and wages}

There are 6 weekly work hours available for an agent: 8, 16, 24, 32, 40 or 48 hours per week. 
Working 8, 16 or 24 hours per week is considered part time work, while working 32, 40 or 48 hours per week is considered full time work. 

The paid wage paid aims to capture the observed distribution of wages, which reflect a variety of job functions.
It depends on the potential wage, weekly hours worked, and wage reduction
\begin{equation}
w^{paid}_{t,g}=h_tw^{pot}_{t,g}(1-r_t),
\end{equation}
where $h_t$ is hours worked divided by 40 hours representing full time working. 

Wage reduction $r_t$ determines how much the offered wage is reduced relative to the potential wage. 
Each unemployment or out of workforce spell increases wage reduction by 4.5-5 percent points annually, while sick leave increases wage reduction by 25 percent points annually (Table \ref{table:reduction}). 
Working either full time or part time reduces wage reduction by 2.5-3 percent points annually. An uninterrupted employment history yields a higher paid wage than a one with significant gaps.

Potential wage $w^{pot}_{t,g}$ is modeled as a simple random autoregressive process (Tanskanen 2020a)
\begin{equation}
w^{pot}_{t,g}=A_{t,g}e^{c\ln(w_{t-1,g}^{pot}/A_{t-1,g})+s-\sigma^2/2},
\end{equation}
where $A_{t,g}$ is socioeconomic group's average wage at age $a_t$ and $s$ is $N(0,\sigma)$ distributed shock term with standard deviation $\sigma=0.05$. 
The relative logarithmic potential wage has $c=89$ percent autocorrelation on the previous year's relative logarithmic potential wage. Finally, term $\sigma^2/2$ is a normalization that keeps the average of the log-normally distributed variable at the desired level.

\begin{table}
\centering
\begin{tabular}{ |l|c|c| } 
 \hline
State & Wage reduction (\%) & Recovery (\%)\\
 \hline
        Full time employed & 0 & 3\\
        Part time employed & 0 & 2.5\\
        Mother's leave & 0 & 0\\
        Father's leave & 0 & 0\\
        Child home care allowance & 2.5 & 0\\
        Disabled & 5 & 0\\
        Retired & 10 & 0\\
        Retired working part time or full time & 5 & 0\\        
        Earnings-related unemployed & 4.5 & 0\\
        Labor market support & 5 & 0\\
        Outside of workforce & 5 & 0\\
        Student or in the army & 0 & 2\\ 
        Sick leave & 25 & 0\\
 \hline
\end{tabular}
 \caption{Annual wage reduction and recovery in employment states}
\label{table:reduction}
\end{table}

Each group corresponds roughly to an education level and gender. Men with a higher education have the highest average wage, while women with a lower education have the lowest average wage.
The observed distribution of wages (Statistics Finland 2024d) is used to fit the variance of wages at various ages.
The average potential wage of each group corresponds to wage distributions at 15, 50 and 85 percentage points of the gender.

\subsubsection{Job search friction}

An agent makes the choice whether to search for a job or not, but the search does not necessarily instantly yield a job. In the LCM, there is an age- and gender-dependent probability for obtaining a job when searching.
These job search probabilities are summarize in Table \ref{table:reduction}, which shows how probable it is for an agent seeking employment to be employed in 3 months. 
Younger agents have a higher probability of receiving a job than older agents. There is a higher probability of obtaining a part time job than a full time job. 
The probabilities were calibrated so that observed employment rates can be reproduced in the model.
The probabilities should not be considered in isolation, since they are conditional on seeking job and hence depend on utility parameters of the respective groups.

Job search friction reduces the employment rate, especially for older workers. 
When an agent searches for a full time job, she or he may still end up with a part time job.
A friction is present also in transitions from part time work to full time work or from full time work to part time work. 

\begin{table}
\centering
\begin{tabular}{ |l|c|c|c|c|c|c| } 
 \hline
full time & men low & men mid & men high & women low & women mid & women high  \\
 \hline
        18-24 & 0.15 & 0.10 & 0.10 & 0.15 & 0.10 & 0.10\\
        25-29 & 0.20 & 0.20 & 0.20 & 0.15 & 0.15 & 0.20\\
        30-49 & 0.25 & 0.25 & 0.30 & 0.30 & 0.30 & 0.30\\
        50-53 & 0.20 & 0.20 & 0.25 & 0.25 & 0.25 & 0.25\\
        55-59 & 0.20 & 0.20 & 0.25 & 0.20 & 0.20 & 0.20\\
        60-64 & 0.10 & 0.15 & 0.20 & 0.15 & 0.20 & 0.20\\
        65-$r_{max}$ & 0.15 & 0.15 & 0.20 & 0.10 & 0.15 & 0.20\\
        $r_{max}$-75 & 0.05 & 0.05 & 0.05 & 0.03 & 0.03 & 0.03\\
 \hline
part time & men low & men mid & men high & women low & women mid & women high  \\
 \hline
18-24 & 0.70 & 0.70 & 0.70 & 0.70 & 0.70 & 0.70\\
25-29 & 0.70 & 0.70 & 0.70 & 0.70 & 0.70 & 0.70\\
30-49 & 0.60 & 0.60 & 0.60 & 0.65 & 0.65 & 0.65\\
50-54 & 0.50 & 0.50 & 0.55 & 0.55 & 0.55 & 0.55\\
55-59 & 0.45 & 0.50 & 0.50 & 0.50 & 0.50 & 0.50\\
60-64 & 0.30 & 0.35 & 0.40 & 0.30 & 0.40 & 0.45\\
65-$r_{max}$ & 0.25 & 0.30 & 0.30 & 0.30 & 0.30 & 0.30\\
$r_{max}$-75 & 0.05 & 0.10 & 0.10 & 0.10 & 0.10 & 0.10\\
 \hline
\end{tabular}
 \caption{Job search friction. Probability of being employed during 3 month period to part time or full time work for various ages, groups and genders.}
\label{table:table5}
\end{table}

\subsection{Taxes and social security benefits}
Wages, taxes, social security benefits and social security contributions influence net income. 
Finnish social security consists of a large number of benefits covering most stages of life, which requires that the Finnish social security is comprehensively implemented in the LCM. 
Still, not every benefit is implemented. In the LCM, benefits are automatically awarded. 


\subsubsection{Taxes and contributions}
Social insurance contributions and income taxes reduce the net income, while VAT reduces means for consumption. 
Municipal taxes, state income taxes, and compulsory public broadcasting company YLE tax are included, but voluntary religious taxes are not. 
VAT is implemented in a simplified way: VAT is paid from all net income that goes over rent. VAT level is the same for all consumption except for housing. 

Social security contributions are modeled in detail. Employer's contributions include health insurance contribution, earnings-related pension insurance contribution, unemployment insurance contribution, and occupational accident and disease insurance contribution. Employee's social security contributions include health insurance contribution, contribution for medical care coverage, contribution for daily allowance, unemployment insurance contribution, employee pension insurance contribution.

Parents of small children pay for day care of their children. While this is not strictly a tax or a social security contribution, it is included.
Wealth, investments or savings are not included, and therefore capital taxes are not taken into account. 

\subsubsection{Unemployment benefits}
Unemployment benefits come in several varieties in Finland: earnings-related benefit, basic unemployment allowance and labor market support. The earnings-related benefit has a higher benefit level than basic unemployment allowance or labor market support. 

The maximum duration of receiving earnings-related benefit is 300, 400 or 500 days depending on the length of person's work history and age. Receiving earnings-related unemployment benefit requires that the person seeks for a job and that she or he is a member of an unemployment fund. The benefits are paid for 5 weekdays each week. The benefit level is computed based on the average wage during the insured employment period required for eligibility. 
Adjusted earnings-related benefits are not included. The extended earnings-related benefit lasting until the retirement is only available for certain older persons. 

Basic unemployment allowance is available to those who are not part of the unemployment fund system, but who would otherwise be eligible for earnings-related unemployment benefit.
The maximum duration of the basic unemployment allowance is the same as in earnings-related unemployment benefit. Labor market support is a means tested unemployment benefit available for non-retired adults. The level of the benefit is the same for Labor market support and for Basic unemployment allowance, and they are here considered a single benefit.

\subsubsection{Pensions}
Finnish pension scheme is an earning related pension scheme that has features from a collective defined contribution scheme, e.g., the life expectancy coefficient. There are four benefits: old-age pension, disability pension, partial early old-age pension, and survivors' pension. All are included in the LCM.

The minimum retirement age varies depending on the cohort (Kannisto and Vidlund 2022). For those born in 1959, the lowest retirement age is 64 years and 3 months. Hence, they would retire at the earliest in 2023. Similarly, the age at which insurance obligation ends also depends on the cohort. For those born in 1959, it is 68 years. In the simulation, the minimum retirement age is uniform and depends on the simulation year.
A retired person may choose to work during retirement with no penalty on the retirement benefit. 

Partial early old-age pension is a separate old-age benefit available to those aged 61 or more. The benefit is 25 or 50 percent of the full retirement benefit. The benefit reduces the full retirement benefit in an almost actuarially neutral fashion.

In addition to an earnings-related pension, there are a means-tested guaranteed pension and a basic pension paid by the state. The accrued earnings-related pension reduces the amount of basic pension by 50 percent. No basic pension is paid, when the accrued earnings related pension is €1,527 per month or more in 2023.

\subsubsection{Other benefits}
Each household pays rent for their apartment in the LCM. Depending on the income, a household may be eligible for one of the two housing benefits in Finland: the general housing benefit and the housing benefit for the retired. 

Supplementary benefit is a guaranteed benefit that ensures that all persons receive a minimum income. Almost all other benefits will reduce supplementary benefit. 
Supplementary benefit has the lowest incentive to work of all social security benefits: above €150 per month, wage reduces the reduction of supplementary benefit fully.

Students receive student benefit automatically. The same level of benefit is paid to all student. This should not influence the results much, since transitions to and from student state occur at external probabilities without the possibility for deciding otherwise. 

Sick leave benefit covers short-term sick leaves.
The maximum duration of sick leave is one year, after which a portion of the agents in sick leave will be moved to disability pension.
There are three states of parental leaves included: Mother's leave, Father's leave and child home care allowance. 
For simplicity, we assume that social security scheme takes full responsibility for child maintenance allowance.

\section{Solving and calibrating}

Our stochastic LCM is based on the Markov decision process framework, in which each decision is made based on the current state without considering the previous states. The chosen action influences, but does not necessarily determine the next transition.
The aim is to find for each state an action that maximizes the expected discounted future utility of an agent. Solving the LCM yields a recipe for choosing an action at each state at each age. This recipe would, at least in principle, tell the optimal action a real person would take in a similar state.

\subsection{Optimization problem}

As in Wu\etal (2017), we consider an agent interacting with a discounted Markov decision process
$(X , A, \gamma, P, u)$.
At each time $t$, an agent chooses the action that maximizes the present value of future utilities. In a policy gradient method, action $a_t\in A$ in the current state $S_t\in X$ is chosen according to policy $\pi(a|S_t)$. 
An action results a transition to the next state $S_{t+1}$ according to the transition probability $P(S_{t+1}|S_t, a_t)$. The agent aims at maximizing the expected $\gamma$-discounted cumulative return 
\begin{equation}
\label{eq:1}
J(\theta)=\max_{a\in A}E[\sum_{s=t}^T\gamma^su(a_s,c_s,S_s,g,h_s)],
\end{equation}
with respect to the policy parameters $\theta$.
Here $t$ is the analyzed time, $A$ is the set of available actions, $u(n,S)$ is the utility, $c_s$ is the consumption at time $s$ in employment state $S_s$, discount factor $\gamma$, $h_s$ is the weekly worked hours, $g$ is group, $a_s$ is agent's age, and $T$ is the length of life cycle.  
To solve the model, we need to find the optimal policy $\pi$.

The current state describes all variables needed for the next decision. After the decision is made, the state changes, and the agent receives a corresponding income.

Reinforced learning is a model-free method of solving Markov decision processes (Sutton and Barto 2018). The two main kinds of deep reinforced learning algorithms are value-based algorithms, such as Q-learning (Mnih\etal 2015), and policy gradient methods used here (Sutton and Barto 2018). The main difference between the methods is in whether the aim is to estimate the value of {\em (state,action)} pairs or more directly estimate the action selection policy. Action policy refers here to the rule connecting a state to an action.

Actor-Critic algorithms are a special case of policy gradient methods. 
Here we use an Actor-Critic algorithm known as ACKTR (Wu\etal 2017), which uses a Kronecker-factored approximation to natural policy gradient allowing the covariance matrix of the gradient to be inverted efficiently (Wu\etal 2017). In our tests, ACKTR converged more rapidly than other Actor-Critic algorithms. Rapid convergence is a desirable feature for an LCMs with a huge state space, even though we are even more interested in obtaining behavior that reproduces observed behavior.

The results of Wang\etal (2019) suggest that Actor-Critic algorithms converge towards the globally optimal policy.
Tanskanen (2022) found that ACKTR approximated dynamic programming well in the LCM setting, which suggests that ACKTR converges toward the global optimum.

For a large state space, only a relatively small portion of the entire state space is encountered in training. This requires that the used method is capable of generalizing the action policy learned from the observed states to those not observed (Sutton and Barto 2018). Neural networks are capable of this kind of generalization, which is the reason reinforced learning methods often use them in estimation.

To improve convergence of the results, we have applied the usual repertoire of tricks: one-hot encoding of employment states, normalization of wages, ages, times in the state and pensions, and scaling of rewards. The policy and value networks are artificial neural nets of with two hidden layers. The neural network is of size (256,256,128) without the output layer. No regularization was applied to the networks in training. Leaky ReLU was used as the activation function for neural networks.

\subsection{Implementation}
The LCM is implemented as three freely available Python modules: {\em Benefits} (Tanskanen 2019a) {\em Econogym} (Tanskanen 2019b) and {\em Life cycle} (Tanskanen 2019c) modules.

{\em Benefits} module (Tanskanen 2019a) describes the Finnish social security scheme, taxation and social security contributions. 
Given parameters describing the state of an agent, it computes the benefits the agent is entitled to, taxes and social security contributions the agent has to pay, net income the agent receives, and consumption available.
The module covers state tax, municipal tax, VAT, and the social security contributions both for an agent, her partner and the household she belongs to as discussed in Chapter 2. 
Benefit module is unit tested against publicly available computations of taxation by, e.g., Veronmaksajat (2023) and Tax administration (2023).

{\em Econogym} module (Tanskanen 2019b) describes how an agent operates during her or his life cycle. 
This modules is compatible with OpenAI Gym environment (Brockman\etal 2016). Given the chosen action, Econogym module advances the state of an agent one time step forward. Econogym module depends on Benefits module.  

{\em Life cycle} (Tanskanen 2019c) is the main module that runs the simulation and keeps track of the results. 
In the training, Life cycle module uses ACKTR to solve the optimal action policy $\pi$, and then uses this policy to guide the simulation to produce aggregate results.
In other words, {\em Life cycle} module trains the LCM, runs the simulation and compiles the results including the statistics. The results can then be analyzed using, e.g., Jupyter notebooks.

\subsection{Calibration}
To produce realistic behavior, we must tune parameters of the LCM appropriately. This is called {\em the calibration step}.
The main data source used in the calibration of the LCM was Employment statistics (Statistics Finland 2024). To fit the wage process, we used Structure of earnings statistic (Statistics Finland 2024d). 
While the use of individual data could be incorporated in the calibration, no register data on individual persons is used in the calibration.

The main targets of calibration are aggregate ratios:
(1) age- and gender-dependent employment rates;
(2) age- and gender-dependent unemployment rates;
(3) age- and gender-dependent proportions of part time workers to full time workers;
(4) age- and gender-dependent disability rates;
(5) aggregate sums of salaries, taxes and individual benefit expenditure;
(6) the proportion of outside of work.

Parts (1), (2), (3) and (6) are taken from Employment statistics (Statistics Finland 2024). Part (4) is from Finnish Centre for Pensions (2024). Part (5) is from Kelasto database (Kela 2024), Finnish Centre for Pensions (2024) or Tax administration (2024). In the calibration, we fit parameters $\kappa$ and re-employment rates in such a way that the result reproduces the observed values as closely as possible.

\subsection{Simulation}
Once the model is calibrated, behavior of a population of agents must be simulated. Simulation produces, e.g., the aggregate benefit expenditure, wage sum and employment rate. 
With a population of 100,000 agents, it takes around 40 minutes on a Macbook Pro 2024 computer to produce the aggregate simulation results. 

\begin{figure}
\includegraphics[height=6cm]{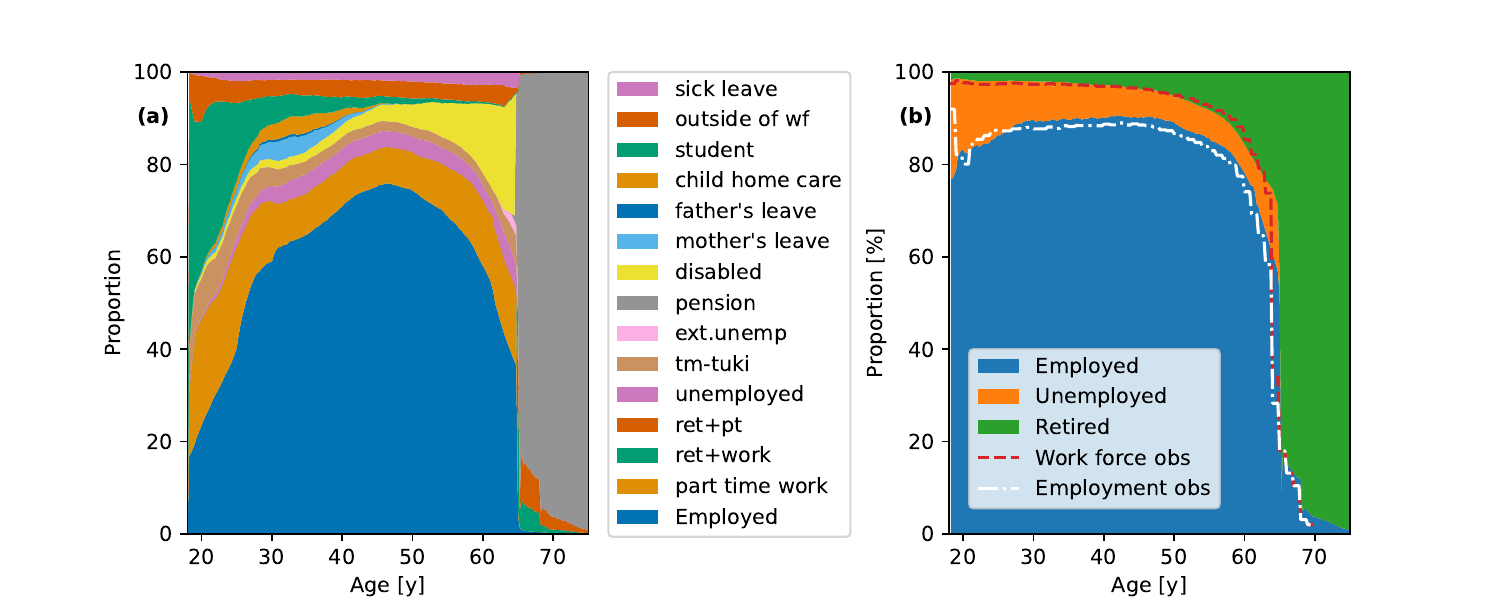}
\caption{(A) Division of the simulated population to employment states in 2023; (B) Division of work force into employed, unemployed and retired compared against Employment statistics in 2023.}
\label{fig:division}
\end{figure}

To understand how stable the results are, we repeat the simulations multiple times. The fitted baseline is used as the starting point. We then make a short refit protocol with 5,000,000 time steps, followed by a population simulation run of a population of 50,000 agents. The size of the simulated population is chosen to roughly correspond to size of the actual cohorts in Finland. The results are mostly averaged over 50 repeats with this protocol. 
This protocol yields a more stable result than training and simulating the model only once.

\section{Baseline results}

Starting from incentives of an individual agent, the LCM aims to replicate observed patterns in employment and unemployment rates, unemployment spells durations, benefit expenditures, aggregate wages, 
participation tax rates and effective marginal tax rates. Baseline results describe how well the LCM captures observed features.

\subsection{Population}

A heterogeneous agent population is essential for generating realistic results.
Fig. \ref{fig:division}A illustrates the distribution of agents in 15 employment states. Most agents under the age of 30 are students.
Between ages 18 and 30, there is a transition from education to employment or unemployment. A relatively high proportion of the workforce takes parental leave between ages 25 and 40. After age 50, a substantial share of the population transitions to disability pension.

Fig. \ref{fig:division}B shows division of work force to employed, unemployed and retired. 
At the statutory minimum retirement age (64 in years 2023\footnote{The statutory minimum retirement age depends on birth cohort. However, for simplicity, we assume it is fixed for each calendar year. This approximation introduces a minor discrepancy.}), there is a rapid transition from employment to retirement. Overall, the model’s distribution aligns closely with observed data.

\subsection{Employment rate}
\begin{figure}
\includegraphics[width=15cm]{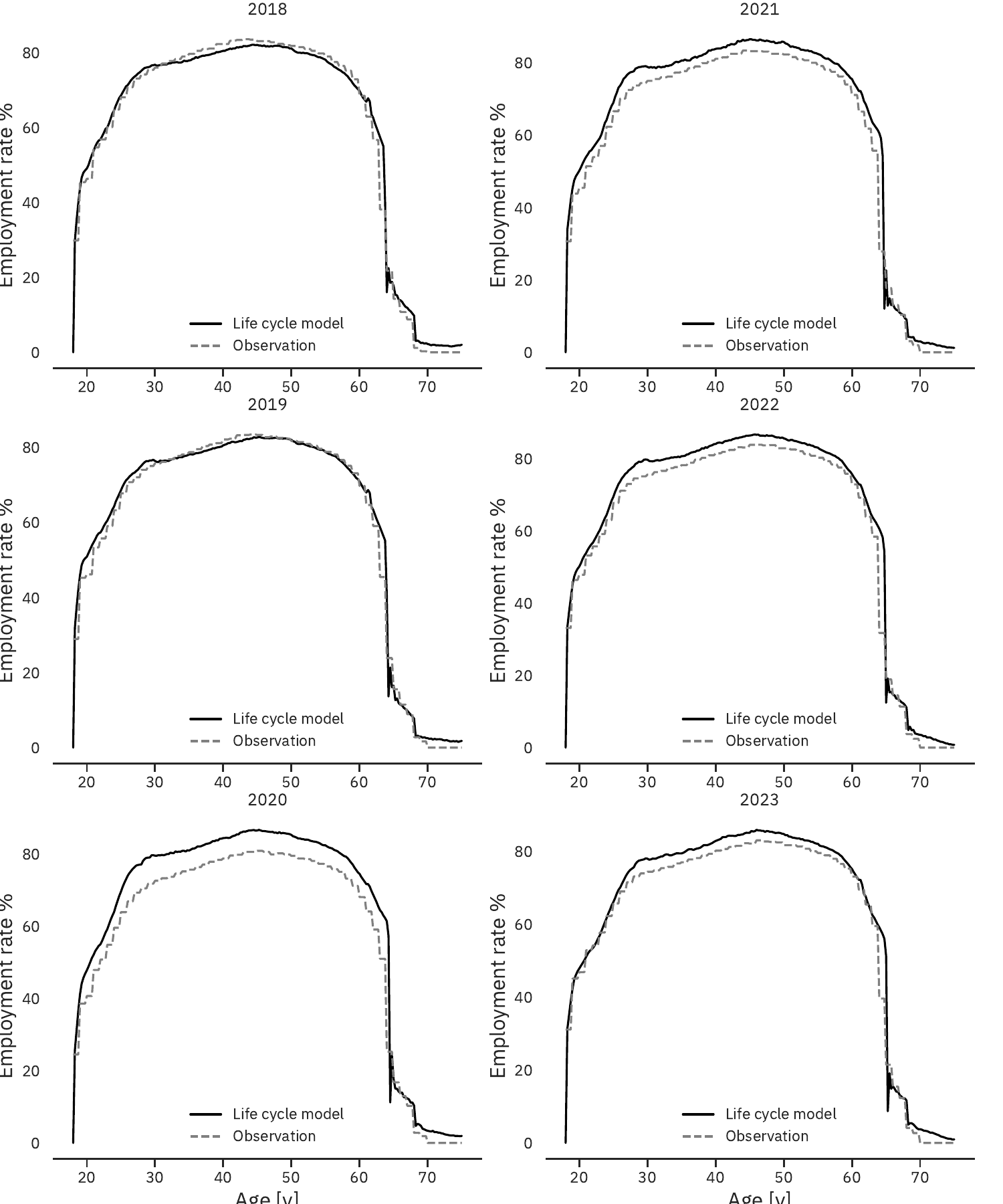}
\caption{Employment rates in 2018, 2019, 2020, 2021, 2022, 2023}
\label{fig:employment_rate}
\end{figure}

Fig. \ref{fig:employment_rate} compares predicted employment rates from 2018 to 2023 against Employment Statistics (Statistics Finland 2024a). The agents' utility parameters remain constant across all years in the simulations, only the initial population conditions, benefit rules, and wage levels vary. 

The model closely matches observed employment rates for 2018 and 2019. The largest deviation occurs in 2020, which is unsurprising given the impact of the COVID-19 pandemic—an event not accounted for in the LCM.
The prediction overestimates employment rates  in years 2021 2022 and 2023. This suggests that job-finding rates (Table \ref{table:table5}) may be too high for these years.

\subsection{Full time and part time work}
Each agent chooses their weekly working hours from a discrete set ranging from 8 to 48 hours, in 8-hour increments. Each choice is associated with a specific penalty term, and the relative magnitudes of these penalties influence the agent’s decision. Part time work is defined as 8–32 hours per week, while full time work is defined as 36 hours or more per week.

\begin{figure}
\includegraphics[width=15cm]{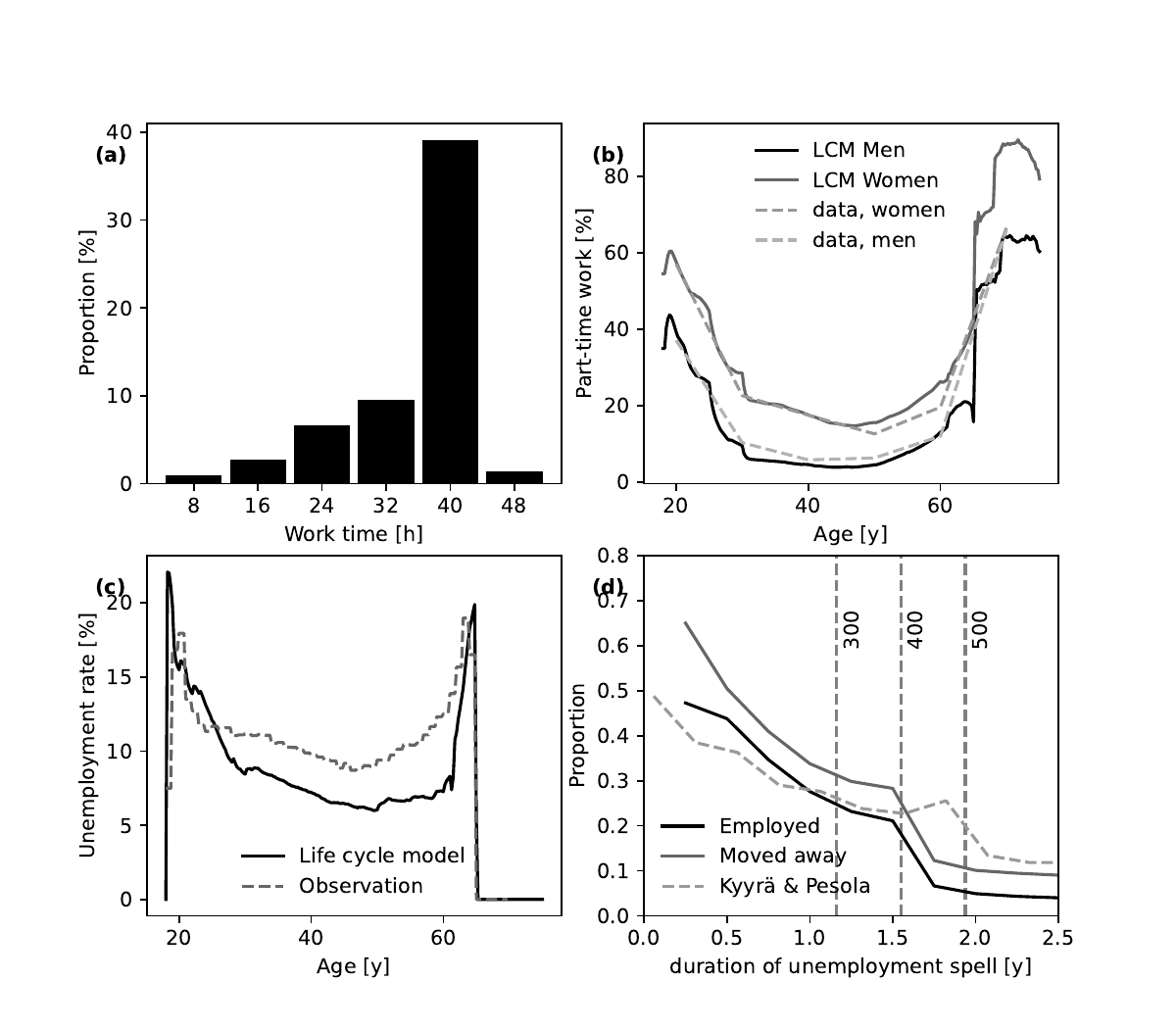}
\caption{
(a) Distribution of weekly work time; (b) Proportion of part time employment in men and women;
(c) Unemployment rate in LCM (solid curve) and observed in 2023 (dashed curve); (d) Duration of unemployment spells in the LCM in 2023.}
\label{fig:parttimework}
\end{figure}

The most common choice among agents is 40 hours per week (Fig. \ref{fig:parttimework}A), with only a small number opting for longer hours. 
These patterns are broadly consistent with the findings of Harju~\etal (2018), however, prevalence of part time work has increased since 2015.
Women are more likely to work part time than men, although this gender difference narrows with age  (Fig. \ref{fig:parttimework}B). 
Part time work is more prevalent among individuals under 30 and over 50, compared to those aged 30–50. 

\subsection{Wages}

\begin{figure}
\includegraphics[width=15cm]{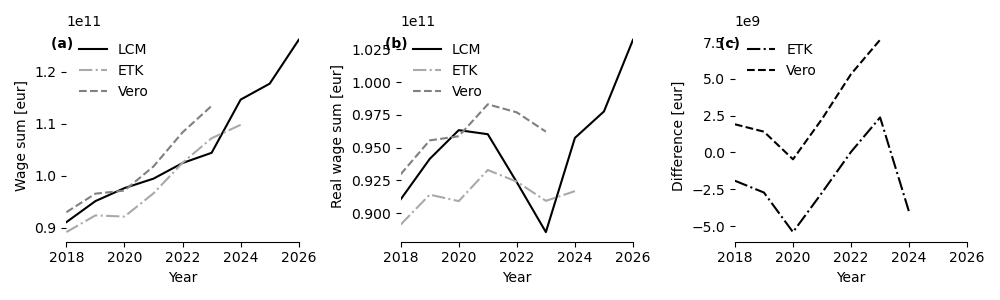}
\caption{Comparison of the LCM results against those from Finnish Tax Authority (Vero) and Finnish Centre of Pensions (ETK) in (A) sum of wages; (B) sum of wages in 2018 money; and (C) difference in sums of wages in 2018 money}
\label{fig:wagefig}
\end{figure}

Sum of wages grows at a roughly same rate as has been observed (Fig. \ref{fig:wagefig}). Real values of wage sum has a large dip in 2023 due to large peak in inflation. Wage sum then recovers in 2024 and continues upward.
Fig. \ref{fig:wagefig} shows that Finnish Tax authority (2024) has higher estimates for the sum of wages including the entrepreneurial income than Finnish Centre for Pensions (2024). 
The predicted sum of wages is in range given by Tax Authority and Finnish Centre for Pensions (Fig. \ref{baseline:wages}).

The sum of wages is underestimated (Table \ref{baseline:wages}) in 2023, while consumption is available to the population is overestimated by €2.5 billion. 

\begin{table}
\centering
\begin{tabular}{|r|r|r|r|r|}
\hline
 & {\bf LCM} & {\bf Observation} & {\bf Difference} \\
\hline
Sum of wages& 104,354,392,658 & 107,206,000,000 & -2,851,607,342 \\
Aggregate benefits & 47,492,714,651 & 47,649,007,900 & -156,293,249 \\
Net income & 94,511,298,269 & 92,010,308,087 & 2,500,990,182 \\
\hline
State tax & 17,404,536,324 & 20,158,338,771 & -2,753,802,447 \\
Municipal tax & 8,345,622,876 & 9,067,601,906 & -721,979,030 \\
Employees pension premium & 7,858,199,207 & 7,601,900,000 & 256,299,207 \\
Unemployment insurance premium & 1,546,709,628 & 1,608,090,000 & -61,380,372 \\
Pension premium & 26,898,596,758 & 26,870,600,000 & 27,996,758 \\
Health insurance premium & 2,340,080,853 & 2,436,042,200 & -95,961,347 \\
Daycare fee & 319,861,881 & 175,800,000 & 144,061,881 \\
VAT & 19,049,625,905 & 21,275,000,000 & -2,225,374,095 \\
YLE tax & 547,995,590 & 558,390,319 & -10,394,729 \\
Employer contribution & 22,818,026,565 & 22,609,745,400 & 208,281,165 \\
\hline
Unemployment benefit, total & 3,689,330,718 & 3,786,095,941 & -96,765,223 \\
Unemployment benefit, ER & 2,441,429,341 & 1,855,124,067 & 586,305,274 \\
Unemployment benefit, non-ER  & 1,247,901,378 & 1,930,971,874 & -683,070,496 \\
Housing benefit & 1,883,490,166 & 2,363,421,555 & -479,931,389 \\
Pensions, total & 33,481,612,923 & 34,026,900,000 & -545,287,077 \\
Pensions, Basic pension & 1,485,455,935 & 2,283,909,667 & -798,453,732 \\
Pensions, Guarantee pension & 428,560,380 & 292,298,054 & 136,262,326 \\
Child maintenance allowance & 819,052,204 & 221,191,667 & 597,860,537 \\
Child benefit & 1,523,272,395 & 1,352,354,007 & 170,918,388 \\
Student allowance & 723,866,524 & 496,626,810 & 227,239,714 \\
Paternity leave & 153,300,293 & 193,579,475 & -40,279,182 \\
Maternity leave & 814,175,674 & 797,086,438 & 17,089,236 \\
Child home care allowance & 262,967,609 & 216,022,447 & 46,945,162 \\
Sickness benefit & 1,346,735,860 & 901,058,982 & 445,676,878 \\
Social assistance & 880,893,970 & 718,462,857 & 162,431,113 \\
\hline
Public finances, net & 32,418,082,298 & 37,666,100,696 & -5,248,018,398 \\
\hline
\end{tabular}
\caption{Wages, taxes, social security benefits and contributions in the baseline LCM. All values are in euros per year.}
\label{baseline:wages}
\end{table}

\subsection{Benefits}
The aggregate error in benefit expenditure is €156 million, that is, 0.33 percent of total expenditure (Table \ref{baseline:wages}). There are estimation errors in both directions in individual benefits.

\subsubsection{Unemployment}
Finland provides three types of unemployment benefits: labor market support, basic unemployment allowance, and earnings-related unemployment benefits. 
The first two offer identical benefit levels and are therefore treated as a single benefit in the model.

The LCM reproduces well the aggregate unemployment benefit expenditure in 2023 well. The prediction error is €97 million, or 2.6 percent of total expenditure (Table \ref{baseline:wages}). However, 
the LCM underestimates unemployment rate (Fig. \ref{fig:parttimework}C). This discrepancy stems from an overestimation of the share of earnings-related unemployment benefits relative to labor market support. Additionally, some of the underestimation arises from increased unemployment benefits not captured in the model.

According to Kela (2023), individuals born outside Finland account for approximately 26 percent of basic unemployment benefit recipients. This demographic group alone may explain up to €500 million of the €683 million difference between observed and predicted basic unemployment benefit expenditure.

The LCM accurately captures the age-dependent pattern of unemployment rates (Fig. \ref{fig:parttimework}C): unemployment is high among the young, declines steadily between ages 20 and 50, and rises again as individuals approach retirement age. A notable spike in unemployment near the minimum retirement age is partly driven by extended earnings-related unemployment benefits.

Fig. \ref{fig:parttimework}D show that the proportion of unemployed that move to any other states is higher than the proportion of those that find a job. 
In other words, the job-finding hazard is lower than the indeterminate exit hazard. This is consistent with findings of Korpela (2023).
Kyyrä and Pesola (2019) find that the likelihood of re-employment decreases with the duration of the unemployment spell but increases sharply at the end of the earnings-related benefit period. 
Korpela (2023) shows that the spike may be due to delay in starting a job, and the spike is likely significantly lower. 

The LCM predicts a transition to employment that quite closely matches the level estimated based on Kyyrä~\etal (2019). 
This calibration of the LCM does not predict that there would be a spike at the end of unemployment, however, if the probability of re-employment is increased, a spike appears at the end of maximum unemployment benefit duration.
Korpela (2023) shows re-employment rates that are lower than those in Kyyrä~\etal (2019), which suggests that the LCM may have too high re-employment rates (Table \ref{table:table5}).

\subsubsection{Other benefits}
Pension expenditure is reasonably accurate with most error coming from underestimation of basic pension expenditure.  
This likely reflects shorter actual working careers than those assumed in the LCM.  The discrepancy may also be partly due to the absence of immigration in the model. Furthermore, because pension benefits are long-term by nature, observed expenditures reflect a mix of regulatory frameworks—some of which differ from those applied in the LCM. Therefore, exact replication of pension expenditure is not expected.

The LCM underestimates housing benefit expenditure by €480 million.  This provides a good target for improved calibration.
Sickness benefit expenditure is overestimated, likely because the real-world system exhibits stronger selection effects than those modeled in the LCM.
Parental leave benefits are predicted well, while alimony payments are overestimated.

Child benefit expenditure is overestimated. This is partly due to the model's current implementation, which—due to technical simplifications—pays child benefits until age 18. This alone accounts for roughly half of the overestimation. The remaining discrepancy is mostly due to the model assigning child benefits to too many households. 

Student allowance expenditure is only roughly approximated in the LCM. Student allowance benefit is paid to students of all ages without checking for eligibility.
The model includes students of various ages and determines transitions in and out of student status via exogenous probabilities. As a result, student allowance expenditure does not significantly influence agents’ work incentives or results.

\subsection{Incentives}

\begin{figure}
\centering
\includegraphics[width=12cm]{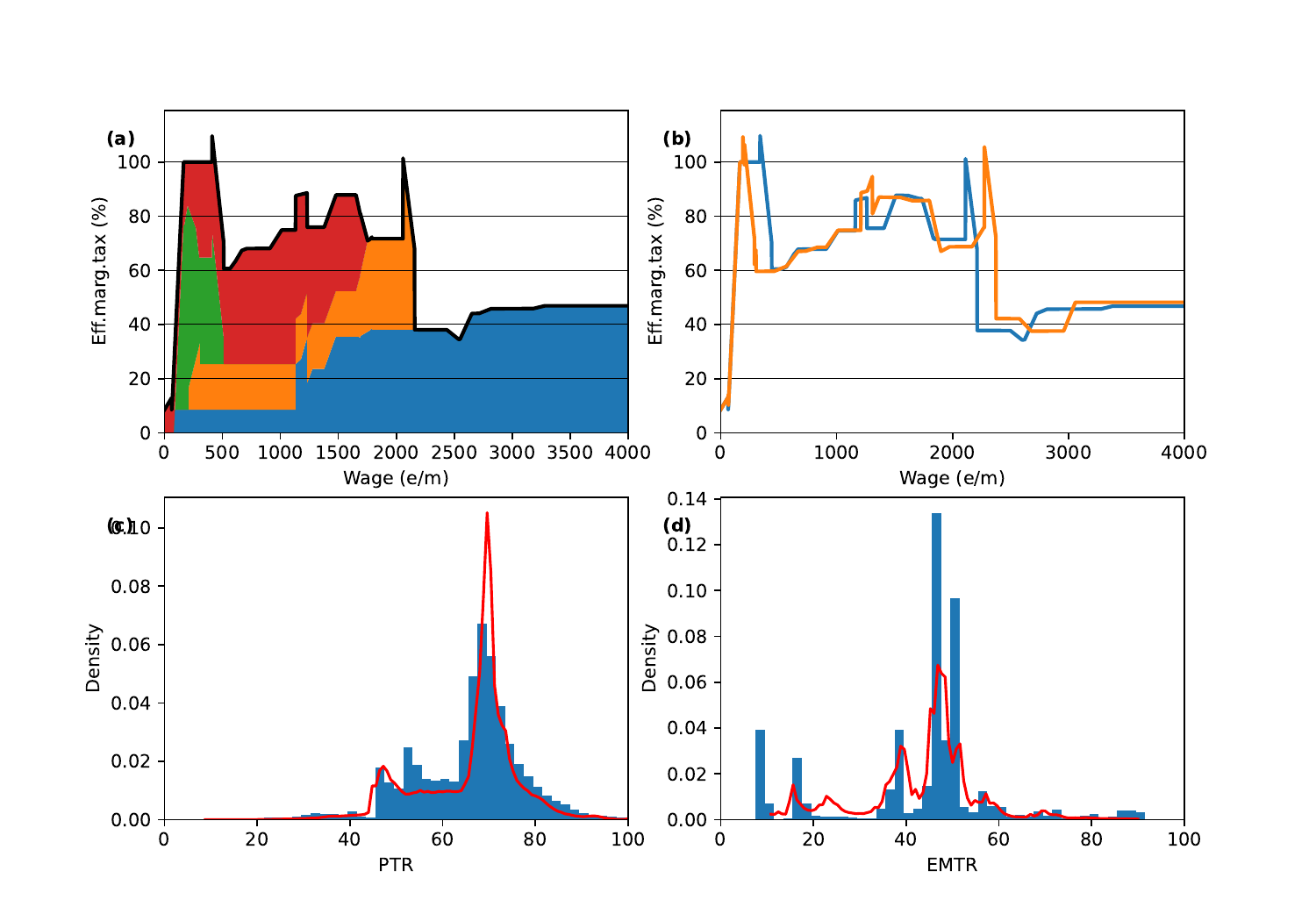}
\caption{Effective marginal tax rate for a single adult with no children and living alone. (A) Composition of effective marginal tax rate. Green is social assistance, red labormarket support, orange housing support and blue taxes and contributions. (B) Comparison of EMTR in years 2022 and 2023; 
(C) Participation tax rate (PTR) in population; (D) Effective marginal tax rate (EMTR). Solid red curve is the EMTR by Puonti \etal (2022). Blue bars are computed using the LCM.}
\label{fig:incentives}
\end{figure}

Fig. \ref{fig:incentives}A compares effective marginal tax rates (EMTR) in years 2022 and 2023 for a single adult living alone receiving Labormarket support.  Fig. \ref{fig:incentives}B shows how various benefits and taxes influenced EMTR in 2023. 

Microsimulation model Sisu (Statistics Finland 2018) can be used to estimate participation tax rates (PTRs) and EMTRs for the 800,000 persons. 
Figures \ref{fig:incentives}C and \ref{fig:incentives}D compares EMTR and PTR estimated by Puonti \etal (2022) against estimation made with the LCM. 
The agreement between the results is good: according to Puonti \etal (2022), the median PTR is 69 percent, while the LCM suggests it is 68.1 percent in 2021. Puonti \etal (2022) estimate that median EMTR is 46 percent, while the LCM estimates 46.9 percent. 
The LCM used here is not based on register data, while Sisu estimate is based on register data. 

Individual spikes in Fig. \ref{fig:incentives}D  correspond to various employment states in the LCM. For example, the highest spike in Fig. \ref{fig:incentives}D largely corresponds to the full time employed state. The difference in the spikes shows that the population in LCM does not fully reproduce the observed population.

\subsection{Public finances}
Overall the LCM underestimates public sector net income (Table \ref{baseline:wages}). 
State tax is underestimated, as can be expected, since state tax is progressive and the LCM does not fully capture all sources of income, e.g., income from business activities. Another reason for the discrepancy is that the wage sum in the LCM is based on the Finnish Centre for Pensions (2024) Statistics database, in which entrepreneur income is based on YEL statistics. This underestimates income of high earning individuals. Similar issues are present in the municipal tax, even though state tax concentrates on high incomes and municipal tax is more of a level tax.

In the social security contributions, there are rather small errors. VAT is rather crudely modeled in LCM, and a more detailed VAT description would yield a more accurate estimate.

\section{Analyzing a reform: Program of Prime Minister Petteri Orpo's Government}

The baseline LCM can be retrained to evaluate the effects of a social security reform or a taxation reform on key outcomes such as public finances or employment.
Retraining the model generates a new optimal action policy, which reflects the revised behavioral responses of individuals under the reformed policy environment. This allows for a direct comparison between the reform scenario and the baseline. Importantly, the preferences of agents and other structural parameters of the LCM remain unchanged during retraining; only the optimal action policy is updated in response to the new policy rules.

Program of Prime Minister Petteri Orpo's Government includes a number of reforms to the Finnish social security scheme. To assess the impact of the program, the most important parts of the reform are implemented as a modification of the LCM. 

\begin{table}
\centering
\begin{tabular}{|r|r|r|r|r|}
\hline
Reform & MoF estimate & Implemented in LCM\\
\hline
Grading of earnings-related UB & 15,800 persons & Implemented\\
The employment condition for UB & & \\ 
increased to 12 months & 5,700 persons & Implemented\\
Abolition of age-related exemptions in UB & 3,900 persons & Implemented\\
Abolition of earnings disregards in UB & - & Implemented\\
The waiting period for UB & & \\increased from five to seven days & 1,000 persons & Not implemented\\
Restoring the periodization of holiday  & & \\
compensation in UB & 2,200 persons & Not implemented\\
Changing the employment condition to  & & \\
earnings-based instead of hours-based & 1,500 persons & Not implemented\\
Removal of the possibility to accumulate  & & \\
work requirement in pay-subsidised work & 1,300 persons & Not implemented\\
\hline
Reductions in personal income taxation & 8,700 persons & Implemented\\
Reform of housing subsidies & 1,900 persons & Implemented\\
\hline
Language requirement for labour market support & 1,300 persons & Not implemented\\
Abolition of job alternation leave & - & Not implemented\\
Abolition of aikuiskoulutustuki & xx & Not implemented\\
Freezing of KEL index & xx & Not implemented\\
\hline
Total & 44,100 persons & 40,000 persons\\
\hline
\end{tabular}
\caption{Reforms of Orpo's government. MoF estimate stands for Ministry of Finance estimates (2024). UB stands for Unemployment benefit.}
\label{reform:mof}
\end{table}

Several of the reforms change unemployment benefits. For example, grading of unemployment benefit means that the initial level of unemployment benefit stays the same for the first eight weeks of benefit (40 benefit
days). Thereafter, the benefit level is 80\% of the initial benefit up to 34 weeks of benefit (170 benefit days), after which the level is 75\% of the initial benefit level.
Table \ref{reform:mof} summarizes the impact of the reforms on employment. We next discuss, how the LCM estimates the impacts.

\subsection{Employment}
\begin{figure}
\includegraphics[height=12cm,clip]{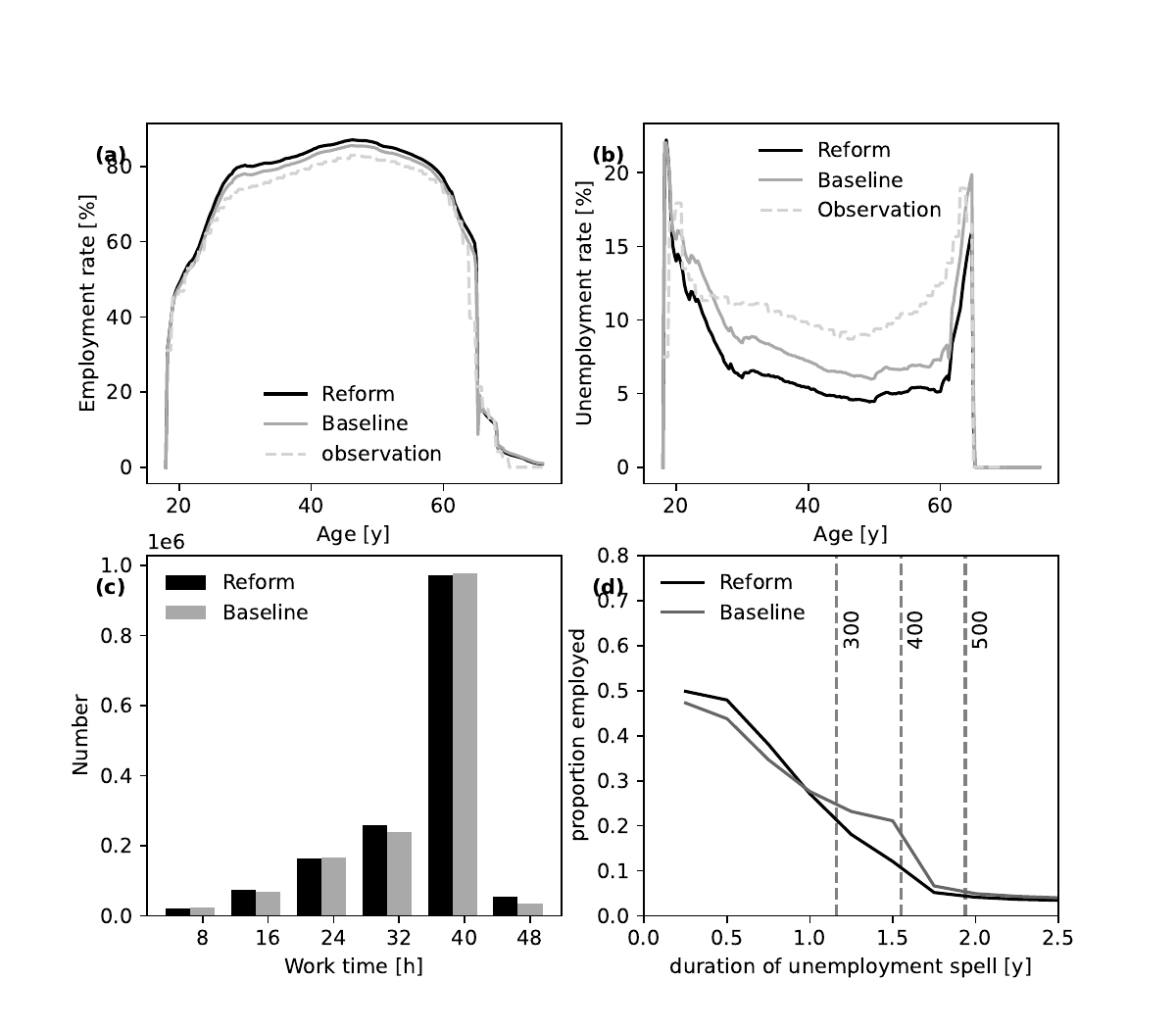}
\caption{(A) Employment rates in the baseline model and the reformed model. (B) Comparison of unemployment rates.
(C) Weekly work hours rate. Y axis refers to the number of agents.
(D) Unemployment spell durations.}
\label{reform:employment_comparison}
\end{figure}

The reform increases employment by 57,865 full time equivalent (FTE; Table \ref{reform:employment_comparison}). This is comparable corresponds to the estimate by the Ministry of Finance (2024) of 44,100 new employed (roughly 40,000 FTE).
The reform increases the employment almost uniformly at all ages (Fig. \ref{reform:employment_comparison}A). As discussed above, an LCM is likely to give a higher estimate on the impact of a reform than elasticity based approach due to a larger number of available impact pathways.

Most of the total increase in employment is due to an increase in the full time employment (Table \ref{table:reform_employment}).
Part time employment and employment during retirement are unchanged.  Interestingly, there seems to be a slight shift from work time of 40 hours per week to both 32 hours per week and 48 hours per week (Fig.\ref{reform:employment_comparison}C).

It is important to understand whether the difference between Reform and Baseline models is significant. In a simulation with 50 iterations, average employment was 2,181,760 FTE in Baseline model and 2,123,895 FTE in Reform model. There is difference of 57,865 FTE between the results (Table \ref{table:reform_employment}). Standard deviation of total employment is 4,998 FTE in the reform model, and 4,082 FTE in Baseline model. To have a significant difference at 99 percent likelihood with 50 iterations, it suffices to have a difference of 2,123 FTE between models. Hence, the difference between employment in the reform and Baseline models is significant.

\subsection{Wages and public finances}
The reform increases annual sum of wages by €3.4 billion (Table \ref{table:reform_financial})
The average wage is €49,395 per year in the Reform and €49,134 per year in the Baseline. The difference is due to shortened and reduced unemployment, which increases the wage offers. 
Sum of wages increases by €2.858 billion due to increased employment and by €556 million due to increase average wages.

The policy of the previous Finnish governments has been to reduce PTR to increase part time working, as evidenced by the introduction of protected portion in several benefits (Kyyrä~\etal 2018). 
Orpo's Government seems to value full time employment more than part time employment, as evidenced by removal of protected portion.
Nevertheless, the LCM predicts that working part time is largely unchanged (Table \ref{table:reform_employment}).
Sum of wages in part time employment is increased from €8.4 billion to €8.8 billion, that is, by €388 million. 

Consumption available to households increases from €95.5 billion to €96.7 billion due to reform, mostly via increased employment. 
In other words, the consumption increases by €1.267 billion. 

\begin{table}[h!]
\centering
\begin{tabular}{rrrr}
\hline
{} & {\bf Reform} & {\bf Baseline} & {\bf Difference} \\
\hline
Employed, total & 2,181,760 & 2,123,895 & 57,865 \\
Employed, part time & 213,049 & 212,450 & 599 \\
Employed, full time & 1,968,711 & 1,911,445 & 57,266 \\
Employed, 18-62 years& 2,080,082 & 2,028,915 & 51,167 \\
Employed, 63+ years& 101,678 & 94,981 & 6,697 \\
Employed, retired & 29,111 & 29,000 & 111 \\
\hline
\end{tabular}
\caption{The impact of the reform on employment measures in FTEs.}
\label{table:reform_employment}
\end{table}

\begin{table}[h!]
\centering
\begin{tabular}{lrrr}
\hline
{} & {\bf Reform} & {\bf Baseline} & {\bf Difference} \\
\hline
Wage sum, total & 107,768,145,926 & 104,354,392,658 & 3,413,753,267 \\
Wage sum, part time & 8,817,050,071 & 8,428,650,757 & 388,399,314 \\
Wage sum, full time & 98,951,095,854 & 95,925,741,901 & 3,025,353,953 \\
Consumption & 95,542,243,977 & 94,511,298,269 & 1,030,945,708 \\
\hline
State tax & 17,923,910,207 & 17,404,536,324 & 519,373,883 \\
Municipal tax & 8,542,042,761 & 8,345,622,876 & 196,419,884 \\
VAT & 19,295,081,838 & 19,049,625,905 & 245,455,934 \\
YLE tax & 557,000,436 & 547,995,590 & 9,004,847 \\
Employees pension premium & 8,110,872,971 & 7,858,199,207 & 252,673,764 \\
Unemployment insurance premium & 1,382,854,384 & 1,546,709,628 & -163,855,244 \\
Pension premium & 27,332,602,843 & 26,898,596,758 & 434,006,085 \\
Health insurance premium & 2,400,228,859 & 2,340,080,853 & 60,148,006 \\
Daycare fee & 334,382,219 & 319,861,881 & 14,520,337 \\
Employer contributions & 23,122,936,755 & 22,818,026,565 & 304,910,189 \\
Public finances, net & 35,421,970,148 & 32,737,944,179 & 2,684,025,968 \\
\hline
\end{tabular}
\caption{The impact of the reform on financial measures. All values are in euros per year.}
\label{table:reform_financial}
\end{table}

LCM predicts that the public finances are strengthened by €2.7 billion. This is due to €1.44 billion increase in taxes and contributions and €1.25 billion reduction in benefit expenditure. 
Ministry of Finance estimates that public finances are strengthened by €1.4 billion.

\subsection{Unemployment}

The reform reduces unemployment rate significantly (Fig. \ref{reform:employment_comparison}B). The underlying reason is the shortening of unemployment spells due to grading of unemployment benefit (Fig. \ref{reform:employment_comparison}D). 
This is particularly obvious near the end of earnings-related unemployment benefit at around 1.5 years, that is, 400 benefit days, after which the insured will receive a lower basic unemployment benefit. 

Another way to look at it is Table \ref{table:unemp_duration}, which shows how unemployment spell durations shorten at various age groups.
There is a significant shift from longer unemployment spells (6m+) to shorter spells (0-6m).
The shift may even be too drastic in young generations, which suggests that re-employment rates may require recalibration.

\begin{table}[h!]
\centering
\begin{tabular}{|c|c|c|c|c|c|}
\hline
\textbf{Reform} \\
\hline
\textbf{Age} & \textbf{0-6 m} & \textbf{6-12 m} & \textbf{12-18 m} & \textbf{18-24 m} & \textbf{over 24 m} \\
\hline
\textbf{20-29} & 0.90 & 0.08 & 0.01 & 0.00 & 0.00 \\
\textbf{30-39} & 0.87 & 0.11 & 0.02 & 0.00 & 0.00 \\
\textbf{40-49} & 0.83 & 0.13 & 0.03 & 0.00 & 0.00 \\
\textbf{50-59} & 0.75 & 0.19 & 0.06 & 0.00 & 0.00 \\
\textbf{60-65} & 0.49 & 0.24 & 0.14 & 0.06 & 0.07 \\
\hline
\textbf{Baseline} \\
\hline
\textbf{Age} & \textbf{0-6 m} & \textbf{6-12 m} & \textbf{12-18 m} & \textbf{18-24 m} & \textbf{over 24 m} \\
\hline
\textbf{20-29} & 0.65 & 0.24 & 0.11 & 0.00 & 0.00 \\
\textbf{30-39} & 0.68 & 0.21 & 0.11 & 0.00 & 0.00 \\
\textbf{40-49} & 0.61 & 0.25 & 0.14 & 0.00 & 0.00 \\
\textbf{50-59} & 0.55 & 0.27 & 0.17 & 0.00 & 0.00 \\
\textbf{60-65} & 0.43 & 0.25 & 0.18 & 0.07 & 0.08 \\
\hline
\end{tabular}
\caption{Distribution of earnings-related unemployment durations binned by age and used earnings-related unemployment benefit days.}
\label{table:unemp_duration}
\end{table}

\subsection{Benefit expenditure}

The reform reduces benefit expenditure by €1,306 million (Table \ref{results:benefits}), of which the majority comes from a reduction of earnings-related unemployment expenditure, €1,117 million. 
The reform reduces basic unemployment benefit expenditure by €188 million as a result of increased employment. Total influence on unemployment benefit expenditure is due to both benefit expenditure reduction and to the decreased unemployment rate.

The reform reduces housing benefit expenditure by €610 million due to reductions in the housing benefit and from increase employment.
The reform has certain other smaller influences, such as an increase of social assistance by €51 million that is due to the reduced unemployment benefit. 
The increase of €674 million in pension expenditure in the long term is due to the increased employment and therefore increase pension accrual. 
The Child benefit is increased in Orpo's program (Table \ref{results:benefits}).

\begin{table}[h!]
\centering
\begin{tabular}{lrrr}
\hline
{} & {\bf Reform} & {\bf Baseline} & {\bf Difference} \\
\hline
Unemployment benefit, total & 2,383,611,118 & 3,689,330,718 & -1,305,719,600 \\
Unemployment benefit, ER & 1,323,636,512 & 2,441,429,341 & -1,117,792,829 \\
Unemployment benefit, basic & 1,059,974,606 & 1,247,901,378 & -187,926,771 \\
Housing benefit & 1,273,683,497 & 1,883,490,166 & -609,806,668 \\
Pension expenditure, total & 34,093,264,043 & 33,481,612,923 & 611,651,120 \\
Pension expenditure, Basic pension & 1,416,214,801 & 1,485,455,935 & -69,241,134 \\
Pension expenditure, Guaranteed pension & 432,548,979 & 428,560,380 & 3,988,599 \\
Alimony & 818,678,168 & 819,052,204 & -374,036 \\
Child benefit & 1,591,226,886 & 1,523,272,395 & 67,954,492 \\
Student allowance & 723,465,220 & 723,866,524 & -401,304 \\
Father leave benefit & 158,598,458 & 153,300,293 & 5,298,165 \\
Mother leave benefit & 815,107,368 & 814,175,674 & 931,693 \\
Child home care allowance & 255,325,848 & 262,967,609 & -7,641,761 \\
Sickness benefit & 1,353,584,578 & 1,346,735,860 & 6,848,718 \\
Social assistance & 932,031,318 & 880,893,970 & 51,137,348 \\
\hline
Sum of benefits & 46,247,340,282 & 47,492,714,651 & -1,245,374,369 \\
\hline
\end{tabular}
\caption{Comparison of LCM estimates of Orpo's reform and the baseline. All values are in euros per year.}
\label{results:benefits}
\end{table}

\section{Discussion}
\subsection{Analyzing a complex social security reform}

Analyzing the impacts of a social security reform is hard. It not only requires that one must compute the static influence of the reform on benefit expenditure, but it must take dynamical effects into account. Dynamic impacts include how the reform changes behavior of people, e.g., the number of people receiving benefit changes, employment rate may change, unemployment rate may change, and even wages may change. These changes may change viability of a reform quite fundamentally. 

Taking various impacts into account requires a complex computational model that mimics behavior of real people. It requires that the model is properly calibrated and incorporates sufficiently large set of results from empirical experiments. 

In this study, we have developed a life cycle model that aims to estimate the impacts of a reform.
The underlying benefit and tax schemes are complex, which required that the state space describing an agent in the model is huge. To solve the LCM, we must find optimal actions of an agent for each of the possible states. This is not an easy task. Nevertheless, a sufficiently optimal action policy could be found using an RL algorithm. After finding the action policy describing behavior of an agent for all states, we can compare how a population of agents behave against the observed employment rates and other measures. The LCM could quite well reproduce the main measures of Employment Statistics (Statistics Finland 2024a).

\subsection{Goodness of calibration}
The LCM was calibrated to reproduce Employment Statistics (Statistics Finland 2024a) successfully, but the LCM reproduces also several elements beyond those explicitly calibrated. These include:
\begin{enumerate}
    \item \textit{Unemployment spell durations}: The LCM approximates the outline of unemployment durations, as reported by Kyyrä~\etal (2019) (Fig. \ref{fig:parttimework}D).  
    \item \textit{Effective Marginal Tax Rates (EMTR) and Participation Tax Rates (PTR)}: The LCM distribution aligns well with results from Puonti ~\etal (2022) (Figures \ref{fig:incentives}C and \ref{fig:incentives}D).  
    \item \textit{Impact analysis}: The LCM's assessment of the Orpo government’s reforms quite closely match findings from the Finnish Ministry of Finance (2024).  
\end{enumerate}  

 These various measures suggest that the estimates made by the LCM are largely in line with other ways of measuring the impacts. 
 This suggests that an LCM could be a valuable tool in the estimation of financial and employment reforms in the future.

\subsection{Interpreting the results}
An LCM produces life trajectories of a single cohort. However, we wish to understand how the current population would respond to financial incentives and reforms.
To better approximate Finland's population in a single year, we scale the size of each cohort to match real-world population structure of Finland. 
This allows us to interpret the results of the LCM simulation as an approximation of the behavior of Finnish population. 

This assumes that each generation lives their lives under the same constant set of rules, or at least behaves as if they did. 
Although these the assumptions are a bit of stretch, a true intergenerational model would require a lot more data and requires an order of magnitude more computing power and complexity in the model. 
Hence, the relatively simplistic assumption that an LCM represents a single year is chosen in this study.

\subsection{Issues and next steps}
The LCM assumes that parameters stay the same for the entire life-time of the cohort under study. If there are large, rapid changes in, e.g., benefits due to price inflation, estimated impacts may be unrealistically large. 

Re-employment rates may need to be revised to better correspond to the observed re-employment rates. 
The dynamics of agents moving into and out of workforce may not accurately capture the actual behavior of people.

There is no migration to or from the population in the LCM. Accuracy of the LCM estimates could be improved by including migration into the model. According to Kela (2023), a significant portion of social security benefits is paid to those born elsewhere. Inclusion of migration may influence the results in either direction since, e.g., employment rate of migrants may not match that of the current population.

It is not fully clear how close to the global optimum the policy is. Since reinforced learning algorithm ACKTR is an approximate gradient descent algorithm, it is possible that ACKTR gets stuck in a local optimum that is not close to the global optimum. There are theoretical studies that suggest that Actor-Critic type algorithms do approach the global optimum in a high-dimensional state space (Fu~\etal 2020). To alleviate this issue, the model was fitted several times, and the results are averaged over several runs. 
Based on the findings of Tanskanen (2022), we believe the model has converged sufficiently close to the global optimum to produce reliable results.

There are many directions to which the LCM could be developed next. Still, one of the most important directions would be to reduce complexity of the model, if possible. 
Inclusion of better models of housing, wealth and savings would be valuable extensions, as would a better model of education. Similarly, a more detailed model of forming couples and having children would improve the results.

\section*{References}

\begin{description}
\item[] Alho, J. M., Kangasharju, A., Lassila, J., Valkonen, T. (2023). {\it Maahanmuutto ja työvoiman riittävyys: Taloudellisten vaikutusten arviointia} (No. 132). ETLA Report.
\item[] Blau, D.M. (2008) "Retirement and consumption in a life cycle model." Journal of Labor Economics 26: 35-71.
\item[] Brockman, G., Cheung, V., Pettersson, L., Schneider, J., Schulman, J. and Zaremba, W. (2016), “OpenAI Gym”, arXiv:1606.01540.
\item[] Browning, M., and Crossley, T. F. (2001). "The life-cycle model of consumption and saving." Journal of Economic Perspectives, 15(3), 3-22.
\item[] Cahuc, P., Carcillo, S., Zylberberg, A. (2014). {\it Labor economics}. MIT press.
\item[] Finnish Centre for Pensions (2024). Statistical Database. https://www.etk.fi/en/research-statistics-and-projections/statistics/statistical-database-2/ (Accessed 2025-Jun-01)
\item[] Fu, Z., Yang, Z., \& Wang, Z. (2020). "Single-timescale actor-critic provably finds globally optimal policy." arXiv preprint arXiv:2008.00483.
\item[] Gourinchas, P. O., \& Parker, J. A. (2002). "Consumption over the life cycle." Econometrica, 70(1), 47-89.
\item[] Hakola, T., Määttänen, N. (2007) ”Vuoden 2005 eläkeuudistuksen vaikutus eläkkeelle siirtymiseen ja eläkkeisiin: arviointia stokastisella elinkaarimallilla.” Eläketurvakeskus.
\item[] Harju, J., Kyyrä, T., Kärkkäinen, O., Matikka, T., and Ojala, L. (2018). {\it Työn tarjonnan mallintaminen osana talouspolitiikan arviointia.} Valtioneuvoston selvitys- ja tutkimustoiminnan julkaisusarja 72/2018. Valtioneuvosto.
\item[] Heer, B., Maussner, A. {\it Dynamic General Equilibrium Modeling: Computational Methods and Applications}, Springer, 2009.
\item[] Hill, A., Raffin, A., Ernestus, M., Gleave, A., Kanervisto, A., Traore, R., Dhariwal, P., Hesse, C., Klimov, O., Nichol, A., Plappert, M., Radford, A., Schulman, J., Sidor, S. and Wu, Y. (2018), “Stable Baselines”, Github repository, https://github.com/hill-a/stable-baselines
\item[] Jiménez‐Martín, S., Sánchez Martín, A. R. (2007). "An evaluation of the life cycle effects of minimum pensions on retirement behavior", Journal of Applied Econometrics, 22(5), 923-950.
\item[] Kannisto, J., and Vidlund, M. (2022). {\it Expected effective retirement age and exit age in the Nordic countries and Estonia.} Finnish Centre for Pensions
\item[] Kela (2024). Kelasto database. https://tietotarjotin.fi/en/statistical-data/2051231/statistical-database-kelasto (accessed 2025-Jun-01).
\item[] Kela (2023). "Maahanmuuttajataustaisten saamissa etuuksissa korostuvat lapsiperhe-etuudet ja ammatillisten taitojen hankkiminen." https://www.kela.fi/ajankohtaista/maahanmuuttajataustaisten-saamissa-etuuksissa-korostuvat-lapsiperhe-etuudet-ja-ammatillisten-taitojen-hankkiminen
\item[] Korpela, H. (2023). "Changing the unemployment insurance duration: heterogeneous effects and an unbudging exit spike." https://www.doria.fi/handle/10024/187918
\item[] Kortelainen, M., Ripatti, A., {\it Estimated DSGE Model of the Finnish Economy: Aino}, Mimeo, Bank of Finland 2010. 
\item[] Kotamäki, M., Mattila, J, Tervola, J. (2018) “Distributional Impacts of Behavioral Effects – Ex-Ante Evaluation of the 2017 Unemployment Insurance Reform in Finland“, Microsimulation 11(2) 146-168
\item[] Kotamäki, M. (2014) "Työllistymisveroasteet Suomessa." Keskustelualoite 1/2014. Valtiovarainministeriö.
\item[] Kyyrä, T., and Pesola, H. (2020). "The effects of UI benefits on unemployment and subsequent outcomes: Evidence from a kinked benefit rule." Oxford Bulletin of Economics and Statistics, 82(5), 1135-1160.
\item[] Kyyrä, T., Matikka, T., and Pesola, H. (2018). {\it Työttömyysturvan suojaosa ja työttömyyden aikainen työskentely.} Valtioneuvoston kanslia 06.07.2018
\item[] Kyyrä, T., Pesola, H., and Verho, J. (2019). "The spike at benefit exhaustion: The role of measurement error in benefit eligibility." Labour Economics, 60, 75-83.
\item[] Mnih, V., Kavukcuoglu, K., Silver, D., Rusu A., Veness J., Bellemare M.G., Graves A., Riedmiller M., Fidjeland A.K., Ostrovski G., Petersen S., Beattie C., Sadik A., Antonoglou I., King H., Kumaran D., Wierstra D., Legg S. and Hassabis D.  (2015) "Human-level control through deep reinforcement learning." Nature 518, 529–533. https://doi.org/10.1038/nature14236
\item[] Ministry of Finance (2024). Monitoring of employment goals (in Finnish), https://vm.fi/tyollisyystavoitteiden-seuranta
\item[] Modigliani, F., and Brumberg, R. (1954). "Utility analysis and the consumption function: An interpretation of cross-section data." In: Kurihara, K.K., Ed., Post-Keynesian Economics, Rutgers University Press, New Brunswick, 388-436.
\item[] Mäenpää, E. (2015). Socio-economic homogamy and its effects on the stability of cohabiting unions. Finnish yearbook of population research, 50.
\item[] Määttänen, N. (2014), “Evaluation of pension policy reforms based on a stochastic life cycle model”, in book Lassila, Määttänen and Valkonen (2014): Linking retirement age to life expectancy – what happens to working lives and income distribution?
\item[] Polvinen, A., Tenhunen, S., Järnefelt, N., Kivelä, M., Kuivalainen, S., Nivalainen, S., Sten-Gahmberg, S. (2023). Työnteko vanhuuseläkkeellä: Kyselytutkimus palkkatyöstä vanhuuseläkkeelle vuosina 2019–2021 siirtyneille.
\item[] Puonti, P., Kauppi, E., Kotamäki, M., and Ropponen, O. (2022). {\it Kannustinloukut Suomessa} (No. 124). ETLA Report.
\item[] Puterman, M.L. (1990) "Markov decision processes." Handbooks in operations research and management science 2: 331-434.
\item[] Rust, J. P. (1989), “A dynamic programming model of retirement behavior”, The economics of aging. University of Chicago Press, 359–404.
\item[] Rust, J. (1997). "Using randomization to break the curse of dimensionality." Econometrica: Journal of the Econometric Society, 487-516.
\item[] Silvo, A., Verona, F. (2020). {\it The aino 3.0 model.} Bank of Finland Research Discussion Paper.
\item[] Statistics Finland (2024a), Employment Statistics (Työssäkäyntitilasto). https://stat.fi/en/statistics/tyokay
\item[] Statistics Finland (2024b), Changes in marital status. https://stat.fi/en/statistics/ssaaty
\item[] Statistics Finland (2024c), Statistics on births. https://stat.fi/en/statistics/synt
\item[] Statistics Finland (2024d), Structure of earnings. https://stat.fi/en/statistics/pra
\item[] Statistics Finland (2024e), Population structure. https://stat.fi/en/statistics/vaerak
\item[] Statistics Finland (2018), Sisu microsimulation model. https://stat.fi/tup/mikrosimulointi/index\_en.html
\item[] Sutton, R.S., and Barto, A.G. (2018). {\it Reinforcement learning: An introduction}, MIT press.
\item[] Tanskanen, A. J. (2019a), Life cycle model, Github repository, https://github.com/ajtanskanen/lifecycle-rl (Accessed 2025-Jun-01)
\item[] Tanskanen, A. J. (2019b), Finnish social security and taxation as a Gym-environment, Github repository, https://github.com/ajtanskanen/econogym (Accessed 2025-Jun-01)
\item[] Tanskanen, A. J. (2019c), Benefits - Python module that makes analysis of social security easy, Github repository, https://github.com/ajtanskanen/benefits (Accessed 2025-Jun-01)
\item[] Tanskanen, A.J. (2020a) "Työllisyysvaikutuksien arviointia tekoälyllä: Unelmoivatko robotit ansiosidonnaisesta sosiaaliturvasta?" Kansantaloudellinen aikakauskirja 2 (2020), 292-321
\item[] Tanskanen, A.J. (2020b) "Kommentti Viherkentälle eläkemaksujen veroluonteesta", Kansantaloudellinen Aikakauskirja 4, 637-641. 
\item[] Tanskanen, A.J. (2022) "Deep reinforced learning enables solving rich discrete-choice life cycle models to analyze social security reforms", Social Sciences \& Humanities Open 5 (1), 100263
\item[] Tanskanen, A.J., Kotamäki, M. (2021) "Lafferin käyrä heterogeenisessa populaatiossa – ja miksi verolajilla on väliä", Kansantaloudellinen Aikakauskirja 117, 383-406. In Finnish.
\item[] Tax Administration (2024) Finnish Taxation database, https://www.vero.fi/en/About-us/statistics/ (Accessed 2025-Jun-01)
\item[] Wang, L., Cai, Q., Yang, Z. and Wang, Z. (2019), “Neural policy gradient methods: Global optimality and rates of convergence”, arXiv preprint arXiv:1909.01150.
\item[] Wu, Y., Mansimov, E., Grosse, R. B., Liao, S. and Ba, J. (2017), “Scalable trust-region method for deep reinforcement learning using kronecker-factored approximation”, Advances in neural information processing systems: 5279–5288.
\end{description}

\end{document}